\documentstyle[12pt]{article}

\advance \topmargin by -\headheight
\advance \topmargin by -\headsep
     
\evensidemargin \oddsidemargin
 \date{}    
\begin{document}
\newcommand{\sect}[1]{\setcounter{equation}{0}\section{#1}}
\renewcommand{\theequation}{\thesection.\arabic{equation}}

\topmargin -.6in
\def\nonu{\nonumber}
\def\rf#1{(\ref{eq:#1})}
\def\lab#1{\label{eq:#1}} 
\def\br{\begin{eqnarray}}
\def\er{\end{eqnarray}}
\def\be{\begin{equation}}
\def\ee{\end{equation}}
\def\0{\nonumber}
\def\lb{\lbrack}
\def\rb{\rbrack}
\def\({\left(}
\def\){\right)}
\def\v{\vert}
\def\bv{\bigm\vert}
\def\lskip{\vskip\baselineskip\vskip-\parskip\noindent}
\relax
\newcommand{\nit}{\noindent}

\newcommand{\ct}[1]{\cite{#1}}
\newcommand{\bi}[1]{\bibitem{#1}}
\def\a{\alpha}
\def\b{\beta}
\def\ca{{\cal A}}
\def\cm{{\cal M}}
\def\cn{{\cal N}}
\def\cf{{\cal F}}
\def\d{\delta} 
\def\D{\Delta}
\def\eps{\epsilon}
\def\g{\gamma}
\def\G{\Gamma}
\def\grad{\nabla}
\def\h{ {1\over 2}  }
\def\hc{\hat{c}}
\def\hd{\hat{d}}
\def\hg{\hat{g}}
\def\hp{ {+{1\over 2}}  }
\def\hm{ {-{1\over 2}}  }
\def\k{\kappa}
\def\l{\lambda}
\def\L{\Lambda}
\def\lg{\langle}
\def\m{\mu}
\def\n{\nu}
\def\o{\over}
\def\om{\omega}
\def\O{\Omega}
\def\p{\phi}
\def\pa{\partial}
\def\pr{\prime}
\def\ra{\rightarrow}
\def\rh{\rho}
\def\rg{\rangle}
\def\s{\sigma}
\def\t{\tau}
\def\th{\theta}
\def\ti{\tilde}
\def\wti{\widetilde}
\def\inte{\int dx }
\def\xb{\bar{x}}
\def\yb{\bar{y}}

\def\tr{\mathop{\rm tr}}
\def\Tr{\mathop{\rm Tr}}
\def\partder#1#2{{\partial #1\over\partial #2}}
\def\ds{{\cal D}_s}
\def\wtwo{{\wti W}_2}
\def\lie{{\cal G}}
\def\alie{{\widehat \lie}}
\def\dlie{{\cal G}^{\ast}}
\def\elie{{\widetilde \lie}}
\def\edlie{{\elie}^{\ast}}
\def\hlie{{\cal H}}
\def\wlie{{\widetilde \lie}}

\def\rlx{\relax\leavevmode}
\def\inbar{\vrule height1.5ex width.4pt depth0pt}
\def\IZ{\rlx\hbox{\sf Z\kern-.4em Z}}
\def\IR{\rlx\hbox{\rm I\kern-.18em R}}
\def\IC{\rlx\hbox{\,$\inbar\kern-.3em{\rm C}$}}
\def\one{\hbox{{1}\kern-.25em\hbox{l}}}

\def\PRL#1#2#3{{\sl Phys. Rev. Lett.} {\bf#1} (#2) #3}
\def\NPB#1#2#3{{\sl Nucl. Phys.} {\bf B#1} (#2) #3}
\def\NPBFS#1#2#3#4{{\sl Nucl. Phys.} {\bf B#2} [FS#1] (#3) #4}
\def\CMP#1#2#3{{\sl Commun. Math. Phys.} {\bf #1} (#2) #3}
\def\PRD#1#2#3{{\sl Phys. Rev.} {\bf D#1} (#2) #3}
\def\PLA#1#2#3{{\sl Phys. Lett.} {\bf #1A} (#2) #3}
\def\PLB#1#2#3{{\sl Phys. Lett.} {\bf #1B} (#2) #3}
\def\JMP#1#2#3{{\sl J. Math. Phys.} {\bf #1} (#2) #3}
\def\PTP#1#2#3{{\sl Prog. Theor. Phys.} {\bf #1} (#2) #3}
\def\SPTP#1#2#3{{\sl Suppl. Prog. Theor. Phys.} {\bf #1} (#2) #3}
\def\AoP#1#2#3{{\sl Ann. of Phys.} {\bf #1} (#2) #3}
\def\PNAS#1#2#3{{\sl Proc. Natl. Acad. Sci. USA} {\bf #1} (#2) #3}
\def\RMP#1#2#3{{\sl Rev. Mod. Phys.} {\bf #1} (#2) #3}
\def\PR#1#2#3{{\sl Phys. Reports} {\bf #1} (#2) #3}
\def\AoM#1#2#3{{\sl Ann. of Math.} {\bf #1} (#2) #3}
\def\UMN#1#2#3{{\sl Usp. Mat. Nauk} {\bf #1} (#2) #3}
\def\FAP#1#2#3{{\sl Funkt. Anal. Prilozheniya} {\bf #1} (#2) #3}
\def\FAaIA#1#2#3{{\sl Functional Analysis and Its Application} {\bf #1} (#2)
#3}
\def\BAMS#1#2#3{{\sl Bull. Am. Math. Soc.} {\bf #1} (#2) #3}
\def\TAMS#1#2#3{{\sl Trans. Am. Math. Soc.} {\bf #1} (#2) #3}
\def\InvM#1#2#3{{\sl Invent. Math.} {\bf #1} (#2) #3}
\def\LMP#1#2#3{{\sl Letters in Math. Phys.} {\bf #1} (#2) #3}
\def\IJMPA#1#2#3{{\sl Int. J. Mod. Phys.} {\bf A#1} (#2) #3}
\def\AdM#1#2#3{{\sl Advances in Math.} {\bf #1} (#2) #3}
\def\RMaP#1#2#3{{\sl Reports on Math. Phys.} {\bf #1} (#2) #3}
\def\IJM#1#2#3{{\sl Ill. J. Math.} {\bf #1} (#2) #3}
\def\APP#1#2#3{{\sl Acta Phys. Polon.} {\bf #1} (#2) #3}
\def\TMP#1#2#3{{\sl Theor. Mat. Phys.} {\bf #1} (#2) #3}
\def\JPA#1#2#3{{\sl J. Physics} {\bf A#1} (#2) #3}
\def\JSM#1#2#3{{\sl J. Soviet Math.} {\bf #1} (#2) #3}
\def\MPLA#1#2#3{{\sl Mod. Phys. Lett.} {\bf A#1} (#2) #3}
\def\JETP#1#2#3{{\sl Sov. Phys. JETP} {\bf #1} (#2) #3}
\def\JETPL#1#2#3{{\sl  Sov. Phys. JETP Lett.} {\bf #1} (#2) #3}
\def\PHSA#1#2#3{{\sl Physica} {\bf A#1} (#2) #3}
\def\PHSD#1#2#3{{\sl Physica} {\bf D#1} (#2) #3}

\begin{titlepage}
\vspace*{-2 cm}
\noindent
\begin{flushright}
\end{flushright}

\vskip 1 cm
\begin{center}
{\Large \bf Bicomplexes and Conservation Laws in Non-Abelian 
Toda Models  } \vglue 1  true cm
 E. P. Gueuvoghlanian  \\

\vspace{1 cm}

{\footnotesize Instituto de F\'\i sica Te\'orica - IFT/UNESP\\
Rua Pamplona 145\\
01405-900, S\~ao Paulo - SP, Brazil}\\
gueuvogh@ift.unesp.br\\ 

\vspace{1 cm}

\end{center}

\normalsize
\vskip 0.2cm

\begin{center}
{\large {\bf ABSTRACT}}\\
\end{center}
\noindent

A bicomplex structure is associated to the Leznov-Saveliev equation of 
integrable models.
The linear problem associated to the zero curvature condition is 
derived in terms of the 
 bicomplex linear equation.  
The explicit example of a Non-Abelian Conformal Affine Toda model is  discussed in detail and its conservation laws 
are derived from the zero curvature representation of its equation of motion.

  \noindent

\vglue 1 true cm

\end{titlepage}

\sect{Introduction}

Two dimensional Toda field theories are examples of relativistic integrable non-linear  
systems underlined by an   Lie  algebraic structure.
Finite dimensional Lie algebras are associated to the Conformal Toda 
Models ( see \cite{balog} for a review ), whose  
 basic representative is   the Liouville model.  The  Affine Toda models are associated
to the loop algebra (centerless Kac-Moody)  
and are characterized by the broken conformal symmetry.   Basic examples within this class, 
we find the 
sine-Gordon,  Lund-Regge (complex 
sine-Gordon), general abelian affine Toda  and homogeneous sine-Gordon \cite{olivetu}. 
Dyonic integrable models such as the singular non abelian Toda models 
are also within  this class \cite{dyonic}.  Conformal Affine Toda 
models \cite{ara254} are related to infinite dimensional Affine Lie 
Algebras (full Kac-Moody algebras). 
Such models are classified according to a grading operator decomposing the Lie
algebra into graded subspaces.  The graded structure is an important ingredient to obtain such models 
 when employing the   hamiltonian reduction procedure  to  the 
WZNW \cite{balog} and 2-loop WZNW models \cite{ara254}. 
Alternatively, the field equations of those models can be obtained from the Leznov-Saveliev 
equation \cite{leznov}. An important fact about this equation is that it can 
be written as a zero curvature condition. As a consequence, 
under specific boundary conditions, there are infinite conserved charges. Also,
if the fundamental Poisson 
bracket relation holds \cite{faddeev}, the involution condition among the 
conserved charges is verified. 
An important ingredient in this construction is the 
classical r-matrix satisfying the classical Yang-Baxter 
equation. 

In recent papers \cite{dimakis1}, a structure called bicomplex was used to 
derive some integrable field equations ( e.g., sine-Gordon, 
non-linear Schroedinger ). It was argued that 
the bicomplex linear equation could, in some cases, lead to chains of 
conserved charges.

In  this paper we generalize the bicomplex structure to derive the    Leznov-Saveliev 
equation corresponding to an infinite dimensional affine Lie algebra, 
which includes the  non abelian Toda equations.
 The linear problem associated to the zero curvature condition is also derived in terms of the
 bicomplex linear equation.  
 Explicit construction, following the arguments of \cite{oli85},
 for  the conserved charges of a specific $A_2^{(1)}$  non-abelian Toda model is obtained.

\sect{Bicomplexes and Leznov-Saveliev equation}

 Let $ V=\oplus_{r\geq0} V^r $
 be an $N_{0}$ -graded linear space over 
$ C $  and $ d, \delta : M^r \rightarrow M^{r+1} $ linear
maps. If $ d^2 = \delta^2 = \delta d + d \delta = 0 $ , then this structure is called a bicomplex \cite{dimakis1}.

It is important to emphasize that nothing is said about Leibnitz rules. Let $ { \xi^1 , \xi^2 }$ 
 be a basis for $ V^1 $ such that $ \xi^1 \xi^1 = \xi^2 \xi^2 = \xi^1 \xi^2 + \xi^2 \xi^1=0 $ .
 In this case $ V^2 $ is one-dimensional and 
$ V = V^0 \oplus V^1 \oplus V^2 $.
It is convenient to introduce light-cone variables in the two-dimensional space-time with 
coordinates $(t,x)$ : $ z=(t+x)/2 $ ; $ \bar{z}=(t-x)/2$ ;
 $\partial = \partial / \partial z= \partial_t + \partial_x $ ;
 $\bar{\partial}= \partial / \partial \bar{z}= \partial_t - \partial_x $ 
.
Consider a infinite dimensional affine Lie 
algebra \cite {god-ol} $ \hat{G}$ and constant generators 
 $( \varepsilon^+ , \varepsilon^- ) \in \hat{G}$  such that
\br
[ \varepsilon^+ , \varepsilon^- ]= \mu_1 \mu_2 \hat{C},
\label {2.1}
\er
where $\hat{C}$ is the central charge generator and $(  \mu_1 , \mu_2 ) \in C $.
 The meaning of 
this choice will be explained in the end of this section. 

Let $ v^1 = ( v^1_1 \xi^1 + v^1_2 \xi^2 ) \in V^1 $ arbitrary  and define:
\br
\delta v^1 \equiv (\delta v^1_1) \xi^1 + (\delta v^1_2) \xi^2; \quad
d v^1 \equiv (dv^1_1) \xi^1 +(dv^1_2) \xi ^2 .
\er
Similarly, for $ v^2 =v^2_{1,2} \xi^1 \xi^2 \in V^2$ arbitrary, define:
\br
\delta v^2 \equiv (\delta v^2_{1,2}) \xi^1 \xi^2 =0; 
\quad d v^2 \equiv (dv^2_{1,2}) \xi^1 \xi^2 =0.
\er

Let $ v^0 \in V^0 $ arbitrary and define  the maps 
$ \delta $, $d$:
\br 
\delta v^0 \equiv \bar{\partial} v^0 \xi^1 + \varepsilon^- v^0 \xi^2; 
\quad  d v^0 \equiv - \varepsilon^+ v^0 \xi^1 + \partial v^0 \xi^2 .
\label{2.2}
\er
An explicit computation reveals that for  $ v^0 \in V^0 $ arbitrary:
\br
\delta^2 v^0 & = &  
                \delta (\bar{\partial} v^0 ) \xi^1 +
                  \delta (\varepsilon^- v^0) \xi ^2 \nonumber \\
             & = & \bar{\partial} (\bar{\partial} v^0 ) (\xi^1)^2 
+ \varepsilon^- \bar{\partial} v^0 \xi^2 \xi^1 
                 + \bar{\partial}(\varepsilon^- v^0) \xi^1 \xi^2 +
(\varepsilon^-)^2 v^0 (\xi^2)^2 \nonumber \\
               &= &  \varepsilon^- \bar{\partial} v^0
 (\xi^2 \xi^1 +\xi^1 \xi^2 )=0 . \nonumber      
\er
That is, 
\br
\delta^2=0; \quad 
d^2=0; \quad 
(\delta d + d \delta ) v^0  = - \mu_1 \mu_2 \hat{C} v^0  \xi^1 \xi^2,
\label{2.5}
\er
where the last two equations are derived 
in a similar way. The last equality can be rewritten as 
\br
P^2 + (\delta d + d \delta) =0, 
\er
where the  map  $ P: V^r \rightarrow V^{r+1}$  is defined by 
\br 
Pv^0 \equiv \varepsilon^+ v^0 \xi^1 + \varepsilon^- v^0 \xi^2. \nonumber
\er
The action of $P$ in $V^1$ and $V^2$ is defined in the same way as it was 
done for $d,\delta$.
Notice that  the maps $(\delta, d)$ do not define a bicomplex, unless the central 
charge is taken equal to zero, which    imply to be working with the
loop algebra. Alternatively, 
let $g$ be an exponential of the generators belonging to $\hat{G}$.  Define a dressing
\cite{dimakis1} for $d$, introducing $D: V^r \rightarrow V^{r+1}$ such that,
for arbitrary $v^0 \in V^0$
\br
Dv^0 \equiv g^{-1} d(gv^0) = -g^{-1} \varepsilon^+ g v^0 \xi^1 +
               ( \partial + g^{-1} \partial g ) v^0 \xi^2.
\er
Extending the action of $D$ in $V^1$ and $V^2$ in the same way as before,
\br
D^2 v^0 = g^{-1} d(gDv^0) 
        = g^{-1} d(g g^{-1}d (gv^0) ) =0 \quad
\rightarrow \quad   D^2 =0,
\label{2.6} 
\er
using (\ref{2.5}) and the fact that $v^0$ is arbitrary.
Now,
\br
(\delta D + D \delta ) v^0 = \{  \bar{\partial}( g^{-1}\partial g )
                    -[g^{-1} \varepsilon^+ g, \varepsilon^-] \}
                      v^0 \xi^1 \xi^2.
\er

In order to get the Leznov-Saveliev equation there are two different options 
here. The first is to take
\br
g=B \exp (- \mu_1 \mu_2 z \bar{z} \hat{C}),
\er
where $B$ is a group element and impose:
\br
\delta D +D \delta = \delta d +d \delta = -P^2.
\er
As a consequence,
\br
\bar{\partial} ( B^{-1} \partial B ) = [ B^{-1} \varepsilon^+ B , 
\varepsilon^- ]; \quad 
\partial ( \bar{\partial} B B^{-1} )= [\varepsilon^+ , B \varepsilon^- B^{-1}].
\label{2.7}
\er 
Equations ( \ref{2.7} ) correspond to the Leznov-Saveliev
equation \cite{leznov} in its two different versions. Notice, however, that the maps 
$(\delta, D)$ defined in this way do not define a bicomplex.

Consider now the second option. Take
\br
g=B  \quad \rightarrow \quad
Dv^0 \equiv B^{-1} d(Bv^0) 
     =      -B^{-1} \varepsilon^+ B v^0 \xi^1 +
               ( \partial + B^{-1} \partial B ) v^0 \xi^2,
\label{2.9}
\er
for $v^0 \in V^0$ arbitrary. Extend the action in $V^1$ and $V^2$ as before
and impose
$
\delta D +D \delta=0.
$ 
This leads to the Leznov-Saveliev equation again and, in this case,
defines a bicomplex:
\br
D^2 = \delta^2 = \delta D + D \delta =0. 
\label{2.11}
\er
An explanation about (\ref{2.1}) is important. 
If $( \varepsilon^+ , \varepsilon^- )$
are choosen in such a way that  (\ref{2.1}) holds,  then 
$ B_0= \exp ( \mu_1 \mu_2 z \bar{z} \hat{C}) $ is a particular 
solution of the Leznov-Saveliev 
equation (\ref{2.7}). 
In fact, this is a vacuum solution required by the dressing method as an input for 
 non-trivial one soliton solutions \cite{dyonic}.

\sect{The Bicomplex Linear Equation}

 A linear problem  is  associated to a given bicomplex in the
following way \cite{dimakis1}: Suppose there is $T^{ (0) } \in V^{0}$ such
that $DJ^{(0)}=0$, where $J^{(0)} \equiv \delta T^{(0)}$. Using (\ref{2.11}),
$ \delta J^{(0)} =0$. Defining $J^{(1)} \equiv D T^{(0)}$ and using 
(\ref{2.11}) , $ \delta J^{(1)} =0$,  $DJ^{(1)}=0$. Suppose that $J^{(1)}$
can also be written as $J^{(1)}=\delta T^{(1)}$,  $T^{(1)} \in V^{(0)}$. Then,
defining $J^{(2)} \equiv DT^{(1)}, \delta J^{(2)} = -D \delta T^{(1)}=
-D J^{(1)}=0$ and $D J^{(2)}=0$. Continuing indefinitely such  steps  and 
defining a formal expansion $T \equiv \sum_{m=0}^{\infty} \rho^m T^{(m)},
 \rho \in C$, the bicomplex linear equation is obtained:
\br
\delta (T- T^{(0)} )= \rho DT
\quad \rightarrow \quad 
\delta T= \rho DT ,
\label{3.1}
\er
if $\delta T^{(0)} =0$.
Using (\ref{2.2}) and (\ref{2.9})  in (\ref{3.1}), result:
\br
\bar{\partial} T = - \bar{A} T
; \quad \partial T = - A T ; \quad
A= - \rho^{-1} \varepsilon^- + B^{-1} \partial B
; \quad \bar{A} = \rho B^{-1} \varepsilon^+ B.
\label{3.5}
\er
Defining
\br
\tilde{\varepsilon^ { \pm} } = \rho^{\pm 1} \varepsilon^ { \pm }
; \quad [ \tilde{\varepsilon^+} , \tilde{\varepsilon^-} ]=
[ \varepsilon^+ , \varepsilon^- ]= \mu_1 \mu_2 \hat{C}, 
\label{3.3}
\er
the commutation relation gets the same structure.
In fact, even the Leznov-Saveliev equation is invariant under 
(\ref{3.3}). Now,
\br
A= -  \tilde{\varepsilon^-} + B^{-1} \partial B
; \quad \bar{A} =  B^{-1} \tilde{\varepsilon^+} B; \quad
\partial \bar{A} - \bar{\partial} A +[A, \bar{A}]=0,
\er
is a standard representation of the connections associated to the
zero curvature equation from which the Leznov-Saveliev equation is derived.
The solution of (\ref{3.5}) is \cite{faddeev}:
\br
T(t,y)= T_0 {\cal P} [ exp (\int^{(t,y)} A_{\mu} dx^{\mu}) ],
\er
where $   \cal{P}$ is the path ordered
operator and $ T_0 $ is a constant.
We should point out that the spectral  parameter $\rho$ naturally arises 
under such framework in (\ref{3.5}).
Given an affine Lie algebra $ \hat{G}$ and a grading operator $ \hat{Q} 
\in \hat{G}$ follows a decomposition \cite{arasup} : 
$
\hat{G} = \oplus_i \hat{ G_{i} }; \quad
[ \hat{Q}, \hat{ G_{i} } ]= i \hat{ G_{i} }; \quad 
[\hat{ G_{i} } , \hat{ G_{j} } ] \in  \hat{ G } _{i+j}. 
$ 
Here, $ i \in Z $. The hamiltonian reduction procedure applied to the
2-loops WZNW model \cite{ara254} leads to the Leznov-Saveliev equation, where
the group element $ B $ is associated to zero-grade generators of $ \hat{G} $
and $ \varepsilon{ \pm } $ are generators of grade $ \pm j , j \in Z $. 
That is, these integrable models are classified in terms of the grading 
operators \cite{arasup}. 
In particular, the class of singular $A_n^{(1)}$ non abelian Toda models
have been constructed in \cite{dyonic} by choosing 
the zero grade subgroup
$\hat{ { \cal G } }_0 = SL(2) \otimes U(1)^{n-1} $. 
 One can realize (\ref{3.3}) as:
\br
\tilde{ \varepsilon^{ \pm  } }=  
exp ( { {\hat{Q} ln \rho  } \o { j } } )    \varepsilon^{ \pm } 
exp ( { {- \hat{Q} ln \rho  } \o { j } } ).
\er 
 
\sect{Conformal Affine Non-Abelian Toda Model}
 
In this section we consider the  example of a $A_2^{(1)}$ conformal
 affine non-abelian Toda model, whose singular version was constructed in 
\cite{dyonic}.  
The zero grade subgroup 
$ \hat{ { \cal G } }_0 = SL(2) \otimes U(1) \subset  {SL}(3) $ is parametrized by
 \br
B=\exp ( \beta \tilde {\chi} E_{-\alpha_{1}}^{(0)})
 \exp (   \beta \varphi_1 H_{\lambda_1}^{(0)} + \beta \varphi_2 h_2^{(0)}+
          \beta \nu \hat{C} + \beta \eta \hat{D} )
 \exp ( \beta \tilde {\psi} E_{\alpha_{1}}^{(0)}),
\er 
\br
\hat{Q} &=& 2 \hat{D} +  H_{\lambda_2}^{(0)}; \quad  
\varepsilon^+ = \mu_1 ( E_{\alpha_{2}}^{(0)} +  E_{-\alpha_{2}}^{(1)} )
; \quad 
\varepsilon^- = \mu_2 ( E_{-\alpha_{2}}^{(0)} +  E_{\alpha_{2}}^{(-1)} ),
\er
where $ \hat{D}$ is the homogeneous grading operator, 
 $h_i^{(0)}= 2 \alpha_i . H^{(0)} / \alpha_i^2  $ are Chevalley generators,
 $H_i^{(0)} $ define the $A_2$ Cartan subalgebra in the Weyl-Cartan basis, 
 $ H_{\lambda_i}^{(0)} = 2 \lambda_i . H^{(0)} / \alpha_i^2 $ ,
 $ \lambda_i $ are the fundamental weights of $A_2$ satisfying
$   2 \alpha_i .\lambda_j  / \alpha_i^2 = \delta_{i,j}$, that is, 
 $ \lambda_i = \sum_{j=1}^2 ( K^{-1})_{i,j} \alpha_j$, $K$ is the Cartan
matrix of $A_2$, $(i,j)=(1,2)$, 
 $\beta^2 = - \beta_0^2 = -(2 \pi)/k$ , $k$ is the WZNW coupling constant. 
Also, the normalization $ \alpha^2 =2 $ for all the roots is adopted
(See  \cite{god-ol} for a description on affine Lie algebras). 

The constant generators $ \varepsilon^{\pm} $ have grade $ \pm 1 $ with respect to the 
generalized grading operator $\hat{Q}$
and  $ B \in \hat{ { \cal G } }_0 $. 
This grading is an intermediate between the
homogeneous grading $ \hat{Q} = \hat{D} $ and the principal grading 
$ \hat{Q} = 3 \hat{D} + H_{\lambda_1}^{(0)} + H_{\lambda_2}^{(0)} $ \cite{arasup}. 
 The Leznov-Saveliev equation leads to  the  field equations 
  corresponding to the lagrangean density
 \br
{\cal L} = (1/3) \partial \varphi_1 \bar{\partial} \varphi_1 +
           \partial \varphi_2 \bar{\partial} \varphi_2+
           (1/2)( \partial \nu \bar{\partial} \eta +
                   \partial \eta \bar{\partial} \nu )+
           \exp ( \beta ( \varphi_1 - \varphi_2 ) )  
              \partial \tilde{\chi}  \bar{\partial} \tilde{\psi} + \nonumber \\
           -( \mu_1 \mu_2 / \beta^2 ) ( \exp (-2 \beta \varphi_2 ) +
                                        \exp (  \beta ( 2 \varphi_2 - \eta  ))
                            + \beta^2 \tilde{\psi} \tilde{\chi} 
                  \exp ( \beta ( \varphi_1 + \varphi_2 -\eta ) ) ).
\label{4.1}
\er

In order to construct the singular non abelian Toda model, 
one observes that $\hat{G}_0^0 \equiv H_{\lambda_1}^{(0)} \in \hat{G}_0 $ 
is such that
$[\hat{G}_0^0,\eps_{\pm} ] =0$ , implying, from the Leznov-Saveliev equation,
 the chiral conservation laws $\partial Tr[ \hat{G_0^0} \bar{\partial} B B^{-1}]=
\bar{\partial} Tr[ \hat{G_0^0}  B^{-1} \partial B]=0$.  Those, in turn allow the 
following subsidiary constraints 
$Tr[ \hat{G_0^0} \bar{\partial} B B^{-1}] = Tr[ \hat{G_0^0}  B^{-1} \partial B] = 0 $, 
responsible for the elimination of the non local
field $\varphi_1$, $
\partial \varphi_1 =
 {3 \o 2} { { \beta \psi \partial \chi \exp (- \beta \varphi_2) } \o {\Delta} }; \quad
\bar{\partial} \varphi_1 =
 {3 \o 2} { { \beta \chi \bar{\partial} \psi \exp (- \beta \varphi_2) } 
\o {\Delta} }, 
$
where
$
\chi = \tilde{\chi} \exp ( \beta \varphi_1 /2 ); 
\psi = \tilde{\psi} \exp ( \beta \varphi_1 /2 );
\Delta = 1+ (3/4) \beta^2 \psi \chi \exp (- \beta \varphi_2 ).
$
The classical r-matrix
associated to this model is discussed in \cite{dubna}. 
The Singular Non-Abelian Affine Toda model, that is, whithout the field 
$ \eta $ ( the field $ \nu $ is only an auxiliary field ) was already 
discussed in the literature. In \cite{dyonic} a complete spectrum of 
1 and 2 soliton solutions was
obtained using the dressing transformations and the vertex operator 
construction. Also, the T-dual
version of this model was analysed and some results about semiclassical 
quantization
were shown.

\sect{Conservation Laws}

In order to derive the  conservation laws for the model defined in (\ref{4.1}),
it is convenient to define a new basis for $ A_2^{(1)} $:
\br
   \hat{Q} &=& 2 \hat{D} +  H_{\lambda_2}^{(0)}; \quad \hat{C}; \quad
A_{(2n)} = \sqrt{3} ( H_{\lambda_1}^{(n)}- (1/6) \delta_{n,0} \hat{C} ) ; \quad
A_{(2n+1)} = E_{ \alpha_{2} }^{(n)} +  E_{ -\alpha_{2} }^{(n+1)}; \nonumber \\
  F_{(2n)} &=& h_2^{(n)} - (1/2) \delta_{n,0} \hat{C}; \quad 
F_{(2n+1)} = E_{ \alpha_{2} }^{(n)} -  E_{ -\alpha_{2} }^{(n+1)}; \nonumber \\
F_{(2n)}^+ &=& E_{ \alpha_{1} }^{(n)}; \quad 
F_{(2n+1)}^+ = E_{ \alpha_{1} + \alpha_{2} }^{(n)}; \quad   
F_{(2n)}^- = E_{ -\alpha_{1} }^{(n)}; \quad
F_{(2n-1)}^- = E_{ -\alpha_{1} - \alpha_{2} }^{(n)}, 
\label{5.1}
\er  
where $n \in Z$.
The generators $ A_{(2n)}, A_{(2n+1)}$ define a infinite dimensional
Heisenberg subalgebra:
\br
[ A_{(2n)}, A_{(2m+1)} ] =0 ;  [ A_{(2n)}, A_{(2m)} ]= 2n \delta_{n+m,0}; 
\quad [ A_{(2n+1)}, A_{(2m+1)} ]= (2n+1) \delta_{n+m+1,0}. 
\label{5.2}
\er
Also, it is verified that
\br
&[& F_{(2n+1)} , A_{(2m)} ] = [ F_{ (2n)} , A_{(2m)} ]=0 ; \nonumber \\  
&[& F_{(2n+p)}^+ , A_{(2m)} ] = - \sqrt{3} F_{ [2(n+m)+p] }^+ \quad ;  
[ F_{(2n-p)}^- , A_{(2m)} ] =   \sqrt{3} F_{ [2(n+m)-p] }^- ; \nonumber \\
&[& F_{(2n+p)}^+ , A_{(2m+1)} ] =   F_{ [2(n+m+p)+1-p] }^+ \quad ;  
[ F_{(2n-p)}^- , A_{(2m+1)} ] = - F_{ [2(n+m+1-p)+p-1] }^- ; \nonumber \\
&[& F_{(2n)} , A_{(2m+1)} ] = 2  F_{ [2(n+m)+1] } \quad ;  
[ F_{(2n+1)} , A_{(2m+1)} ] = 2 F_{ [2(n+m+1)] } , 
\er
where $ p=(0,1) $. Defining
$ { \cal A }= \{  A_{(2n+p)}, \hat{Q} , \hat{C} \} $ and
$ { \cal F }= \{ F_{(2n+p)}, F_{(2n \pm p)}^{ \pm } \} $ ,
we see that linear combinations of generators $ \in { \cal F} $ 
 are used to construct the vertex operators \cite{arasup},
used in the dressing method 
\cite{dyonic}.

The conservation laws follow from the zero-curvature equation by gauge transforming 
$A$ and $\bar A$ into $A^{R}_{ab}$ and $\bar A^{R}_{ab}$ such 
that $[A^{R}_{ab},\bar A^{R}_{ab}]=0$  \cite{oli85},\cite{aramod}. 
 It is convenient to
define the notation:
$
{ \cal F } = { \cal F }^+ \oplus { \cal F }^- \oplus { \cal F }^0 ; \quad
{ \cal A } = { \cal A }^+ \oplus { \cal A }^- \oplus { \cal A }^0 ,
$
where the subspaces     
$ ( { \cal F }^{\pm} , { \cal F }^0 ) $ have a (positive/negative,zero) grade
w.r.t. $ \hat{Q} $ and similarly to $ ( { \cal A }^{\pm} , { \cal A }^0 ) $. 
Consider 
\br
A= B \varepsilon^- B^{-1} ; \quad
\bar{A} = - \varepsilon^+ - \bar{\partial} B B^{-1}; \quad
g_R= [ \prod_{m=1}^{\infty} \exp ( S_{-m} ) ] \exp ( \xi \varepsilon^- ),
\label{7.78}
\er
where the connections result in the Leznov-Saveliev equation under the 
zero-curvature equation and 
$ S_{-m} $ is a linear combination of generators in ${ \cal F }^-$
and $ \xi (z, \bar{z} )$ is a function of $z$ and $\bar z$. Consider  the gauge transformation:
\br
A^R &=& g_R A g_R^{-1} - \partial g_R g_R^{-1}
= \sum_{m=- \infty }^{-1} ( A_{ { \cal A } }^{R,m} +
                               A_{ { \cal F } }^{R,m} ); \nonu \\
\bar{A}^R &=& g_R \bar{A} g_R^{-1} - \bar{\partial} g_R g_R^{-1}
= \sum_{m=- \infty }^1 \bar{A}^{R,m}
\label{5.4}
\er
where $ \bar{A}^{R,m} \in \hat{G}_m $,  
$  A_{ { \cal A } }^{R,m} $ has grade $m$ and is a linear combination
of generators in ${ \cal A }$ and similarly for  $ A_{ { \cal F } }^{R,m} $. 
Explicitely,
\br
\bar{A}^{R,1} = - \varepsilon^+ ; \quad 
\bar{A}^{R,0} = - \bar{\partial} B B^{-1} + \mu_1 \mu_2 \xi \hat{C} 
                - [ S_{-1} , \varepsilon^+ ];
\quad \quad \quad \quad \quad \quad \quad
\nonumber \\
\bar{A}^{R,-1} =
         - \bar{\partial} (S_{-1} +  \xi \varepsilon^-)
        - [ S_{-2}, \varepsilon^+ ] 
        - (1/2) [ S_{-1}, [ S_{-1} , \varepsilon^+ ] ] - 
               [ \xi \varepsilon^-  +  S_{-1} , \bar{\partial} B B^{-1} ];
... 
\er          
\br
\bar{\partial} B B^{-1} = 
\beta [ \bar{\partial} \varphi_1 
          - (3/2) \beta \tilde{\chi} \bar{\partial} \tilde{\psi} 
                  e^{ \beta( \varphi_1 -\varphi_2 ) }  ]  H_{\lambda_1}^{(0)}
          + \beta [ \bar{\partial} \varphi_2 + 
                  (1/2) \beta \tilde{\chi} \bar{\partial} \tilde{\psi} 
                  e^{ \beta( \varphi_1 -\varphi_2 ) } ] h_2^{(0)} + 
\quad \quad \quad
\nonumber \\
+  \beta [ \bar{\partial} \nu \hat{C} + \bar{\partial} \eta \hat{D}] + 
\beta \bar{\partial} \tilde{\psi} e^{  \beta( \varphi_1 -\varphi_2 ) } 
E_{ \alpha_1 }^{(0)}
+ \beta [ \bar{\partial} \tilde{\chi} 
          + \beta \tilde{\chi} \bar{\partial} ( \varphi_1 - \varphi_2 )
          - \beta^2 \tilde{\chi}^2 \bar{\partial} \tilde{\psi} 
                           e^{ \beta( \varphi_1 -\varphi_2 ) } ]
                               E_{- \alpha_1 }^{(0)} 
\quad \quad
\nonumber \\     
\equiv             \bar{J}^{(1)} H_{\lambda_1}^{(0)} +
                   \bar{J}^{(2)} h_2^{(0)} +
                   \bar{J}^{ \nu } \hat{C} +
                   \bar{J}^{ \eta } \hat{D} +
                   \bar{J}^+ E_{ \alpha_1 }^{(0)} +
                   \bar{J}^- E_{- \alpha_1 }^{(0)} .
\quad \quad \quad \quad \quad \quad \quad \quad \quad \quad \quad \quad
\quad \quad 
\label{5.7} 
\er
Now, choose $ \{ S_{-m} \} $ and $ \xi $ such that:
\br
\bar{A}^{R} =  - \varepsilon^+
                   -  \bar{J}^{ \eta } \hat{D}_{ \cal A } 
                   -  \bar{J}^{(1)}  H_{\lambda_1}^{(0)}
                  +  \sum_{m=- \infty }^{-1}
                   \bar{a}^{R,m} A_{(m)}; \nonumber \\
                  \hat{D} \equiv \hat{D}_{ \cal A } + \hat{D}_{ \cal F }
     = [ (1/2) \hat{Q} -   { \sqrt{3}  \o {12} }  A_{(0)} - (1/6) \hat{C} ] 
                  + [ - { F_{(0)} \o 4 } ]. 
\label{5.3}
\er
The idea behind this structure is to solve for $S_{-m}$ such that  all terms  
in 
$ { \cal F } $ are eliminated.
Consider the zero-curvature equation and $ ( A^R , \bar{A}^R ) $ as 
described. Note that 
$ A_{ { \cal A } }^{R,m} = a_{ { \cal A } }^{R,m} A_{(m)}$.
In terms of the new basis (\ref{5.1}), we find
\br
\partial \bar{J}^{ \eta } =0; \quad 
\partial \bar{J}^{(1)} =0; \quad
a_{ { \cal A } }^{R,-1}=0; \quad 
\partial \bar{a}^{R,-1}=0; \nonumber \\
\partial \bar{a}^{R,-m} - \bar{\partial} a_{ { \cal A } }^{R,-m}
- { m \o 2} \bar{J}^{ \eta }  a_{ { \cal A } }^{R,-m} =0, m \geq 2; \quad
A_{ { \cal F } }^{R,m} =0, \forall m.
\label{5.9}
\er
where $\bar {J}^{\eta}$ and $\bar {J}^{(1)}$ are defined in (\ref{5.7}).
 Under a gauge transformation defined by the group element $g_R^R$, follows:
\br
g_R^R= \exp [ - \varepsilon^{+} \int_{ \bar{L} }^{ \bar{z} }
       e^{   - {1 \o 2}  \int_{ \bar{L} }^{ \bar{w} }
        \bar{J}^{ \eta } (\bar{v}) d \bar{v}  } d \bar{w}   ]
         \exp [ -  \int_{ \bar{L} }^{ \bar{z} } 
        (   \bar{J}^{(1)} (\bar{w}) H_{\lambda_1}^{(0)} 
                +  \bar{J}^{ \eta } (\bar{w}) \hat{D}_{ \cal A }  ) 
        d \bar{w}  ];
\er 
\br
\partial \bar{A}^R_{ab} - \bar{\partial} A^R_{ab}=0;
\er
where
\br
A^R_{ab}       &\equiv&  \sum_{m=-\infty}^{-1} A^{R,m}_{ab}= 
                        \sum_{m= -\infty }^{-1} a_{ { \cal A } }^{R,m} 
                         e^{   - {m \o 2}  \int_{ \bar{L} }^{ \bar{z} }
                          \bar{J}^{ \eta } (\bar{v}) d \bar{v}  }  A_{(m)};
 \nonumber \\                                                         
\bar{A}^R_{ab} &\equiv& \sum_{m=-\infty}^{-1} \bar{A}^{R,m}_{ab}
                              + a^{ \hat{C} } \hat{C} \nonumber \\
               &=     &\sum_{m= -\infty }^{-1} \bar{a}^{R,m} 
             e^{   - {m \o 2}  \int_{ \bar{L} }^{ \bar{z} }
        \bar{J}^{ \eta } (\bar{v}) d \bar{v}  }  A_{(m)}
        - \mu_1  \bar{a}^{R,-1} 
         e^{    {1 \o 2}  \int_{ \bar{L} }^{ \bar{z} }
               \bar{J}^{ \eta } (\bar{v}) d \bar{v}  } 
         \int_{ \bar{L} }^{ \bar{z} }
         e^{   - {1 \o 2}  \int_{ \bar{L} }^{ \bar{w} }
               \bar{J}^{ \eta } (\bar{v}) d \bar{v}  }  d \bar{w}  
             \hat{C},
\er
where $ \bar{L} \in R $. 
Taking  
$ \eta(z, \bar{z}) = \eta_1 (z)+ \eta_2 (\bar{z})$ as solution for the equation of motion for $\eta$, results 
$  \int_{ \bar{L} }^{ \bar{z} } \bar{J}^{ \eta } (\bar{v}) d \bar{v}
= \eta_2 (\bar{z})- \eta_2 (\bar{L})$. This implies that all the terms in the
abelian connections are local, except the term in $\hat{C}$.  
Under periodic boundary conditions \cite{oli85},\cite{aramod}, the zero curvature equation
for the abelianized connections implies  an infinite set of 
conserved charges:
\br
\partial_t Q^R_m =0, m \leq -1; Q^R_m =\int_{-s }^{s } A^{R,m}_{x,ab} (t,y) dy;
\quad A^{R,m}_{x,ab}= {1 \o 2} ( A^{R,m}_{ab} - \bar{A}^{R,m}_{ab} ),
\er
where $s \in R$. In order to verify the involution condition, one starts from the 
Fundamental Poisson
Bracket relation (see for instance \cite{dubna}):
\br
\{ A_{x}^{S}(y,t){\otimes }
A_{x}^{S}(z,t)\}_{PB} =[ r,A_{x}^{S}(y,t)\otimes I + 
I \otimes A_{x}^{S }(z,t) ]
\delta (y-z); \quad \quad \quad \quad \nonumber \\
A_{x}^{S} = 
              {1 \o 2 } S(A- \bar{A})S^{-1} -  \partial_x SS^{-1}; \quad 
               S= e^{ - {1 \o 2}  
               ( \beta \varphi_1 H_{\lambda_1}^{(0)} 
               + \beta \varphi_2 h_2^{(0)}+
                \beta \nu \hat{C} + \beta \eta \hat{D} ) }
                e^ {- \beta \tilde {\chi} E_{-\alpha_{1}}^{(0)} }; \nonumber \\
        r=  { \beta^2 \o 4 } (C^+ - \sigma C^+); 
               \sigma (a \otimes  b)=   b \otimes a, \forall (a,b) 
               \in \hat{G}; \quad \quad \quad \quad \quad \quad 
                            \quad \quad \quad \quad \nonumber \\
C^{+}={\sum _{m=1}^{\infty}}
{{\sum _{a,b=1}^{2}}}\frac{\alpha _{b}^{2}}{2}(K^{-1})_{a,b}\left(
h_{a}^{(m)}\otimes h_{b}^{(-m)}\right) +\frac{1}{2}
{
\sum _{\alpha >0}}\frac{\alpha ^{2}}{2}
\left( E_{\alpha }^{(0)}\otimes E_{-\alpha
}^{(0)}\right) + \quad \quad \quad \nonumber \\
+{\sum _{m=1}^{\infty }}{\sum _{\alpha >0}}
\frac{\alpha ^{2}}{2}\left[ E_{\alpha }^{(m)}\otimes E_{-\alpha
}^{(-m)}+E_{-\alpha }^{(m)}\otimes E_{\alpha }^{(-m)}\right], \quad \quad \quad\quad \quad \quad \quad \quad \quad \quad
\er
where $(A, \bar{A})$ are defined in (\ref{7.78}).
 As a consequence \cite{oli85},\cite{aramod},
\br
\{ trT^m,trT^n \} =0, \quad T= {\cal P} [ exp (\int_{-s}^{s} A_x (t,y) dy) ],
\er
$(m,n) \in Z$. Since $trT^n$ are gauge invariant quantities, the previous 
relation holds also for the abelian connections $A^{R}_{x,ab}$. It then follows
that \cite{oli85},\cite{aramod}:
\br
\{ Q^R_m , Q^R_n \} =0, \forall (n,m).
\er
Another set of conservation laws can be obtained in a completely analogous way by considering positive grade expanssion in (\ref{7.78}).
The relevant equations are  summarized in the appendix.

\sect{Conclusion}
In this paper a bicomplex structure associated to the generalized 
Leznov-Saveliev 
equation is established. In this sense, the bicomplex structure is equivalent
to the zero-curvature equation. 
Also, the linear problem associated to the zero curvature condition is derived 
in terms of the
bicomplex linear equation.
The conservation laws for a non-abelian Toda model were obtained, 
generalizing the standard procedure in the abelian models.

{\bf Acknowledgements} The author thanks J. F. Gomes, G. M. Sotkov and A. H. 
Zimerman for discussions and FAPESP for financial support. 

\sect{Appendix}
Let
$A= \varepsilon^- + B^{-1} \partial B; \quad 
\bar{A}= - B^{-1} \varepsilon^+ B; \quad 
g_L= 
[ \prod_{m=1}^{\infty} \exp ( S_{m} ) ] \exp ( \bar{ \xi } \varepsilon^+ ),
$ \\
where $ S_{m} $ is a linear combination of generators in ${ \cal F }^+$
and $ \bar{ \xi } (z, \bar{z} )$ is a function. \\ 
Under a gauge transformation:
$
A^L = g_L A g_L^{-1} - \partial g_L g_L^{-1}
= \sum_{m=-1}^{ \infty }  A^{L,m}; \\
\bar{A}^L = g_L \bar{A} g_L^{-1} - \bar{\partial} g_L g_L^{-1}
= \sum_{m=1 }^{ \infty } ( \bar{A}_{ { \cal A } }^{L,m} +
                               \bar{A}_{ { \cal F } }^{L,m} ),
$ \\
where $ A^{L,m} \in \hat{G}_m $,  
$  \bar{A}_{ { \cal A } }^{L,m} $ has grade $m$ and
is a linear combination
of generators in ${ \cal A }$. Similarly for 
$ \bar{A}_{ { \cal F } }^{L,m} $.
Also, \\  
$
B^{-1} \partial B  = 
\beta [ \partial \varphi_1 
          - (3/2) \beta \tilde{\psi} \partial \tilde{\chi} 
                  e^{ \beta ( \varphi_1 - \varphi_2 ) }  ]
            H_{ \lambda_1}^{(0)}
          + \beta [ \partial \varphi_2 + 
                  (1/2) \beta \tilde{\psi} \partial \tilde{\chi} 
                  e^{ \beta ( \varphi_1 -\varphi_2 ) } ] h_2^{(0)}+ 
\quad \quad \quad
\nonumber \\
+  \beta [ \partial \nu \hat{C} + \partial \eta \hat{D}] + 
\beta \partial \tilde{\chi} e^{  \beta( \varphi_1 -\varphi_2 ) } 
E_{- \alpha_1 }^{(0)}
+ \beta [ \partial \tilde{\psi} 
          + \beta \tilde{\psi} \partial ( \varphi_1 - \varphi_2 )
          - \beta^2 \tilde{\psi}^2 \partial \tilde{\chi} 
                           e^{ \beta( \varphi_1 -\varphi_2 ) } ]
                               E_{ \alpha_1 }^{(0)} 
\quad \quad
\nonumber \\     
\equiv             J^{(1)} H_{ \lambda_1 }^{(0)} +
                   J^{(2)} h_2^{(0)} +
                   J^{ \nu } \hat{C} +
                   J^{ \eta } \hat{D} +
                   J^- E_{- \alpha_1 }^{(0)} +
                   J^+ E_{ \alpha_1 }^{(0)} .
\quad \quad \quad \quad \quad \quad \quad \quad \quad \quad \quad \quad
\quad \quad \\  
$
Solving for $ (S_m , \bar{\xi}) $ such that \\
$
A^{L} =  \varepsilon^-
                   +  J^{ \eta } \hat{D}_{ \cal A } 
                   +  J^{(1)}  H_{\lambda_1}^{(0)}
                  +  \sum_{m=1 }^{ \infty }
                   a^{L,m} A_{(m)}, 
$ \\
the zero curvature equation leads to \\
$
\bar{\partial} J^{ \eta }=0; \quad
\bar{\partial} J^{(1)}=0; \quad
\bar{\partial} a^{L,1}=0, \bar{A}_{ { \cal F } }^{L,m}=0, \forall m. \\
$
$
\partial \bar{a}_{ { \cal A }}^{L,m} - 
\bar{\partial} a^{L,m}
+{ m \o 2} J^{ \eta }  \bar{a}_{ { \cal A } }^{L,m} =0, m \geq 2;
\bar{a}_{ { \cal A }}^{L,1}=0; \\
$ 
where 
$
\bar{A}_{ { \cal A } }^{L,m}= \bar{a}_{ { \cal A}}^{L,m} A_{(m)}.  \quad  \\
$
Under another gauge transformation defined by \\
$
g_L^L= \exp [ \varepsilon^{-} \int_{ L }^{ z }
       e^{   - {1 \o 2}  \int_{ L}^{ w }
        J^{ \eta } (v) d v  } d w   ]
         \exp [ \int_{ L}^{ z}  
        (   J^{(1)} (w) H_{\lambda_1}^{(0)} 
                +  J^{ \eta } (w) \hat{D}_{ \cal A }  ) 
        d w  ]; \\
$ 
follows,
$
\bar{A}^L_{ab}       \equiv  \sum_{m=1}^{\infty} \bar{A}^{L,m}_{ab}= 
                        \sum_{m= 1}^{\infty} \bar{a}_{ { \cal A } }^{L,m} 
                         e^{   {m \o 2}  \int_{ L }^{ z }
                          J^{ \eta } (v) d v  }  A_{(m)};
                           \quad  \partial \bar{A}^L_{ab} - 
                            \bar{\partial} A^L_{ab}=0;
 \nonumber \\                                                         
A^L_{ab} \equiv \sum_{m=1}^{\infty} A^{L,m}_{ab}
                              + b^{ \hat{C} } \hat{C} \nonumber \\
               =     \sum_{m= 1 }^{\infty} a^{L,m} 
             e^{   {m \o 2}  \int_{ L }^{ z }
        J^{ \eta } (v) d v  }  A_{(m)}
        - \mu_2  a^{L,1} 
         e^{    {1 \o 2}  \int_{ L }^{ z }
               J^{ \eta } (v) d v  } 
         \int_{ L }^{ z }
         e^{   - {1 \o 2}  \int_{ L }^{ w }
               J^{ \eta } (v) d v  }  d w  
             \hat{C}, \\
$
where $ L \in R $. \\
The conserved charges are obtained:
$
\partial_t Q^L_m =0, m \geq 1; Q^L_m =\int_{-s }^{s } A^{L,m}_{x,ab} (t,y) dy;
$ where $s \in R$.

\end{document}